\documentclass[10pt,aps,pre,amsmath,amsfonts,amssymb,floatfix,longbibliography,superscriptaddress,notitlepage]{revtex4-1}

\usepackage{mathrsfs} \usepackage{amssymb} \usepackage{graphicx,color} \usepackage{bbm} \usepackage{multirow}
\usepackage{bbm} \usepackage{mathrsfs} \usepackage{commath}\usepackage{tikz}
\usetikzlibrary{shapes,arrows,positioning,calc}\usepackage{stmaryrd}\usepackage{bm}

\usepackage[colorlinks=true]{hyperref}
\hypersetup{
  colorlinks=true,
  linkcolor=blue,
  filecolor=magenta,
  citecolor=blue,
  urlcolor=blue,
}

\def\to{\rightarrow}

\newcommand{\eq}[1]{Eq.~(\ref{#1})}

\newcommand{\beq}{\begin{equation}} \newcommand{\eeq}{\end{equation}}
\newcommand{\beqn}{\begin{eqnarray}} \newcommand{\eeqn}{\end{eqnarray}}

\begin{document}

\title{Orientational ordering of closely packed Janus particles}

\author{Kota Mitsumoto}
\affiliation{Graduate School of Science, Osaka University, Toyonaka, Osaka 560-0043, Japan}

\author{Hajime Yoshino}
\affiliation{Cybermedia Center, Osaka University, Toyonaka, Osaka 560-0043, Japan}
\affiliation{Graduate School of Science, Osaka University, Toyonaka, Osaka 560-0043, Japan}

\begin{abstract}
  We study orientational ordering of $2$-dimensional closely packed Janus particles by extensive Monte Carlo simulations.
  For smaller patch sizes the system remains in the plastic crystal phase
  where the rotational degrees of freedom are disordered down to the
  lowest temperatures. There the liquid consist of dimers and trimers
  of the attractive patches. For large enough patch sizes, the system exhibits a thermodynamic transition
  into a phase with stripe patterns of the patches breaking the three-fold rotational symmetry.
  Our results strongly suggests that the latter is a 2nd order phase transition
  whose universality is the same as that of the $3$-state Potts model in $2$-dimensions. 
  Furthermore we analyzed the relaxation dynamics of the system performing quenching simulations
  into the stripe phase. We found growing domains of the stripes. The relaxation of key
  dynamical quantities follow 
universal scaling features in terms of the domain size.
\end{abstract}

\maketitle

\section{Introduction}

Colloids offer possibilities to create new condensed matters
starting from the design of the building blocks: the size, shape and mutual interactions of the colloidal particles.
The primary interaction between the colloids is the hard-core repulsion.
Oftenly they also have attractive interactions on top of it.
If the attractive interaction is sufficeintly short-ranged, they are called as
{\it sticky colloids} which attract each other if and only if
they are in {\it contact}: their surfaces are separated only by
a negligibly short distance.
Sticky colloids exhibit various interesting phenomena
such as crystal-crystal transitions\cite{bolhuis1994prediction,shin2013self},
glass-glass transitions\cite{fabbian1999ideal,sciortino2002disordered},
gas-liquid critical phenomena\cite{miller2003competition}, gelation\cite{lu2008gelation}, etc. 

Moreover colloids can have rotational degree of freedoms.
If the colloids are non-spherical like ellipsoids, the mutual
interaction becomes dependent not only on the mutual distance between the centers of the particles but also on their
orientations in complex manners. The resultant anisotropic interactions bring about
richer variation of gels \cite{madivala2009self,ruzicka2011observation}.
The so called patchy colloids, which have the sticky attractive interactions but only through their patch like limited regions on their surfaces,
provide varieties of soft condensed matters.
In particular simple spherical colloids with designed patches on top of them offer opportunities to study various non-trivial
consequences of anisotropic interactions in a controlled way.
It is known that patchy colloids realize various stable phases depending on the interaction strength, number and size of the patches.\cite{bianchi2006phase,sciortino2009phase,sciortino2010numerical,romano2011colloidal,chen2011directed,iwashita2013stable,vissers2013predicting} 

\begin{figure}
  \centerline{
 \includegraphics[width=0.8\textwidth]{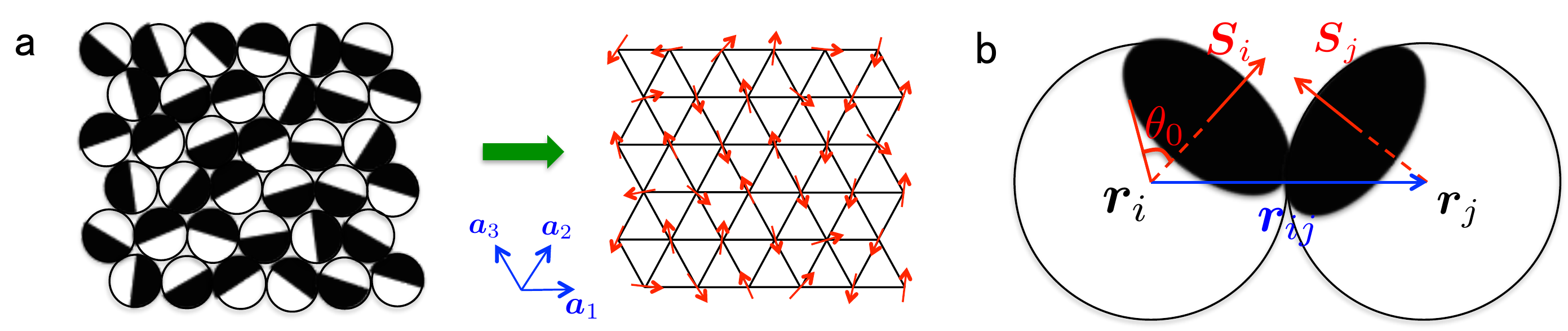}
  }
  \caption{
    (a) (Left) Closely packed Janus particles in 2d. The black region represent
    the patches of the particles. 
    The center of mass of the particles are fixed on the triangular lattice but the particles can rotate.
    (a) (Right) Thus the system can be simply defined as a system of spins (red arrow) on the triangular lattice
    with three ridges being parallel to the following three unit vectors (blue)
    ${\bf a}_{1}=(1,0)$,
    ${\bf a}_{2}=(1/2,\sqrt{3}/2)$,
    ${\bf a}_{3}=(-1/2,\sqrt{3}/2)$.
    (b) Schematic diagram of a pair of Janus particles, $i$ and $j$,
    located at $\bm{r}_{i}$ and $\bm{r}_{j}$ in the closely
    packed configuration. The distance between their centers of mass
    $r_{ij}=|\bm{r}_{j}-\bm{r}_{i}|$
    is the same as their diameter. The patches (black) attract each other
    when they directly touch with each other.
    The surface area of the patch can be specified by the angle $\theta_{0}$.
    The orientation of the $i$-th Janus particle
    can be specified by the 'spin' or 'director' $\bm{S}_i$
    which is a unit vector.
}
\label{fig_janus}
\end{figure}

In the present paper, we study the simplest spherical patchy colloids
with just one patch on each of them, which are called as "Janus particles", put in the $2$-dimensional closest packing configuration, i.e. the triangular lattice of the spherical particles directly touching all the nearest neighbors. (see Fig.~\ref{fig_janus}(a)). In such an configuration particles cannot move but just rotate.
Thus the configuration of the system can be described solely by 'spins' or 'directors' $\bm{S}_i$ ($i=1,2,\ldots,N$)
which represent the orientations of the patches of the particles
$i=1,2,\dots,N$.
This is a very interesting set up which realizes a simplest {\it plastic crystal}: orientationally disordered states on top of a translationally ordered, 
crystalline background.
Indeed such set ups have been implemented and analyzed experimentally.\cite{iwashita2014orientational,jiang2014orientationally}.
However the nature of the orientational ordering have not been fully studied yet. While a theoretical study \cite{shin2014theory} predicts an orientationally ordered state
for large enough patch sizes, an experiment \cite{jiang2014orientationally} suggests an orientational
glass phase. In the present work, we perform extensive Monte Carlo simulation to study both the static and dynamic aspects of the orientational configurations.

The attractive interaction between two Janus particles is switched on
only if their patches are directly touching with each other as
in Fig.~\ref{fig_janus} (a). Let us say that the two particles
have an {\it attractive contact} in such a situation.
Then a key parameter of the system is the patch size
which can be parameterized by the angle $\theta_0$ (see Fig.~\ref{fig_janus} (b)).
The spatial patterns of the directors in the ground states changes dramatically
with the patch size (see Fig.~\ref{fig_equilibrium_data} (a)).
1) {\bf Dimers} ($0^\circ < \theta_0 \le 30^\circ$):
Each particle can have attractive contacts at most with 1 other adjacent
particle so that a ground
state configuration consists of dimers of patches.
This is just a dimer covering problem on the triangular lattice
so that there are energetically degenerate, numerous ground states.

2) {\bf Trimers} ($30^\circ < \theta_0 \le 60^\circ$)
Each particle can have attractive contacts at most with
2 other particles so that a ground
state configuration consists of trimers of patches.
Again energetically degenerate, numerous disordered ground states 
can be expected. 3) {\bf Stripes} ($60^\circ < \theta_0 \le 90^\circ$):
Each particle can have attractive contacts at most with 3 other adjacent
particles. A ground state consists of stripes of the patches.\cite{shin2014theory}

One would naturally expect rotationally disordered
states (plastic crystals) with dimers or trimers
at low enough temperatures in the regime 1) and 2).
A possibility in the regime 3) is a long-ranged ordering of the stripes.
Shin and Schweizer studied and analyzed the system theoretically based on a harmonic approximation from the ground states
and proposed a theoretical phase diagram. \cite{shin2014theory}
They argued that there are dimer liquid, trimer liquid and crystalline stripe phases at low temperatures which are separated from the
high temperature, totally disordered orientational liquid (plastic crystal)
phase by 1st order transitions.
However the validity of the harmonic approximation up to higher temperatures
may be questioned. Thus it is not obvious whether all the transitions are real thermodynamic transitions nor they are all first order transitions.
Rotational ordering of the closely packed Janus particles have also attracted interests in experiments\cite{iwashita2014orientational}
which found qualitative agreements with the theoretical phase diagram
but the precise nature of the transitions remains unknown.
In order to clarify these issues we performed a detailed analysis of the putative orientational phase transitions by equilibrium Monte Carlo simulations. 

Dynamics of the rotational degree of freedom was also studied
experimentally in \cite{jiang2014orientationally}  which reported that Janus particles with patches size $\theta_0 = 90^\circ$, making the stripe patterns,
show glassy dynamics avoiding long-ranged ordering of stripes. 
The possibility of creating an orientational glass from a plastic crystal is a very interesting question by itself \cite{suga1974thermodynamic} which deserves to be investigated thoroughly.
Indeed if the equilibrium transition to the stripe phase is a 1st order transition as suggested theoretically, super-cooling
of the orientationally disordered state and the ultimate orientational
glass transition is not inconceivable.
In order to clarify this issues we performed a detailed analysis of the orientational dynamics after temperature quenches by Monte Carlo simulations. 

In this paper, we report the following two sets of main results.
The first is from the equilibrium Monte Carlo simulations.
We found that the transition to the dimer and trimer states are not thermodynamic phase transitions but smooth crossovers.
Moreover we found that transition to the stripe phase is not a 1st order transition but a 2nd order phase transition.
In the stripe phase there are 3 possible directions of the stripes and the phase transition breaks the
$3$-fold rotational symmetry.
We defined an appropriate order parameter which is able to detect the breaking of the 3-fold rotational symmetry. 
We found that the universality class of the transition is the same as $3$-state Potts model in 2 dimensions.
The second set of results is obtained from the analysis on the relaxational dynamics of the system after temperature quenches by the Monte Carlo simulations.
We found that the stripe phase ordering kinetics obey a universal scaling feature of the standard coarsening phenomena
instead of the glassy dynamics at odds with the experimental suggestion.

The organization of the paper as follows.
In sec. \ref{sec-model} we introduce the model and 
in sec. \ref{sec-MC} we explain our simulation method.
Then in sec. \ref{sec-equilibrium-MC} we report the results
of equilibrium Monte Carlo simulations.
And in sec. \ref{sec-out-of-equilibrium-MC} we report the results
of out-of equilibrium Monte Carlo simulations.
In sec. \ref{sec-implications-for-experiments} we discuss implications
of our results for experiments. Finally in sec .\ref{sec-Conclusion}
we conclude the paper with a few remarks.

\section{Model}
\label{sec-model}

In the present paper we focus on the rotational degree of freedoms of the
closely packed Janus particles in 2 dimension (Fig.~\ref{fig_janus} (a))
freezing out the translational motions for simplicity.
The center of mass position ${\bf r}_{i}$
of the Janus particles $i=1,2,\ldots,N$ are put on the triangular lattice
whose three ridges being parallel to the following three unit vectors
    ${\bf a}_{1}=(1,0)$,
    ${\bf a}_{2}=(1/2,\sqrt{3}/2)$,
    ${\bf a}_{3}=(-1/2,\sqrt{3}/2)$.
Then the system can be considered as an assembly of classical vectors representing the directors $\bm{S}_i$ ($i = 1,2,...N$)
put on the triangular lattice.
For simplicity we describe the directors by
$2$-component spin vectors $\bm{S}_i=(S^{x}_{i},S^{y}_{i})$
which lie in the same plane of the triangular lattice.
Note that here we are disregarding the out-of-plane components
of the directors which we believe to be irrelevant.
The Hamiltonian is given by,
\begin{equation}
H=\sum_{<i,j> }
  V(\hat{\bm{r}}_{ij}\cdot\bm{S}_i,\hat{\bm{r}}_{ij}\cdot\bm{S}_j)
\end{equation}
with 
\begin{equation}
V(x,y)  =- \epsilon \Theta(x - \cos\theta_0) \Theta(-y - \cos\theta_0)
\end{equation}
which is the rotational part from Kern-Frenkel potential. \cite{kern2003fluid}
Here $\hat{\bm{r}}_{ij}=({\bf r}_{j}-{\bf r}_{i})/|{\bf r}_{j}-{\bf r}_{i}|$
and $\Theta(x)$ is the Heaviside's step function : $\Theta(x)=1(x\ge0),-1(x<0)$.
Note that on the triangular lattice, $\hat{\bm{r}}_{ij}$ is parallel to one of the three
vectors ${\bf a}_{1}$,${\bf a}_{2}$,${\bf a}_{3}$ shown in Fig.~\ref{fig_janus} (a).
Patches size is parameterized by the angle $\theta_0$.

It can be easily seen that the potential $V$ takes a constant negative value $-\epsilon$ ($\epsilon>0$) if and only if the two particles are in
an attractive contact, i.e. they are in contact through their patches.
As the result, the energy levels of the system become discretized in spite
of the fact that the spins are continuous variables.
We introduce a dimensionless energy $\tilde{E}=H/\epsilon$ (hereafter we call it simply as $E$)
and a dimensionless temperature $\tilde{T} = k_BT/\epsilon$ (hereafter we call it simply as $T$)
where $k_{\rm B}$ is the Boltzmann's constant.  We denote the thermal average in equilibrium as $\langle \ldots \rangle_{\rm eq}$.

Let us comment on the symmetry of the system. The system is
invariant under global rotations, by which not only the spins but also
the whole lattice are rotated all together. Note that the system is not invariant under
rotations of the spins only at variance with usual spin systems.
Since we have fixed the translational degree of freedoms in the present study,
the global rotational symmetry is anyway lost.
More important is the fact that the spin Hamiltonian
is invariant global rotations by a discrete angle $2\pi/3$
reflecting the triangular lattice. The ground states with the ordered
stripe patterns (see Fig.~\ref{fig_equilibrium_data} (a)) break the 3-fold rotational symmetry.

\section{Monte Carlo simulation}
\label{sec-MC}

We performed Monte Carlo simulations with the Metropolis method.\cite{landau2014guide} 
In order to facilitate the computation using vector computers,
we employed the sub-lattice flipping for spin updates. 
To update a spin $\bm{S}_{i}$, a new candidate spin configuration $\bm{S}'_{i}$ was created by
rotating $\bm{S}_{i}$ by a random angle $\gamma$, which obeys a 
uniform distribution in the range $-\pi/3 \leq \gamma < \pi/3$.
In a unit of Monte Carlo step (MCS) a spin update is tried for all spins $\bm{S}_{i}$ ($i=1,2,\ldots,N$).
In the present work, we used system size $L=48,72,108$ for the
equilibrium simulations and $L=144,288,576,1152$ for the out-of equilibrium simulations.
We employed periodic boundary conditions in both simulations.

\begin{itemize}

  \item{\bf Equilibrium Monte Carlo simulations}

In equilibrium simulations, we observed static physical quantities for various patches sizes
$\theta_{0}$ and different temperature $T$.
The equilibrium states are realized by following slow cooling procedure.
(0) Fully equilibrated configurations are prepared at a high enough temperature $T=0.70$
for each patches size $\theta_{0}$.
(1) The temperature is lowered by a small step $\Delta T=0.01$.
(2) Then the system is thermalized by  $3\times 10^{5}$ (MCS).
(3) Observations of physical quantities are made during additional $3\times 10^{5}$ (MCS) to evaluate their thermal averages $\langle \ldots \rangle_{\rm eq}$
by their time averages. (4) return to (1).

In the vicinity of the transition temperatures, the temperature step
$\Delta T=0.01$ used in the annealing is replaced by a smaller one $\Delta T=0.002$.
To conform that the equilibration time is large enough, we performed
simulations using a longer equilibrating time  $3\times10^{6}$ (MCS)
and checked that the results do not change.

We performed $50$ 
statistically independent runs and evaluated the time averages and
the mean-squared errors of the physical quantities.

\item{\bf Out-of equilibrium Monte Carlo simulations}

To study the relaxation dynamics after temperature quenches into the stripe phase,
we mainly worked with the patches size $\theta_0 = 90 ^\circ$.
The unit of time is $1$ (MCS).
We quenched the system from $T = \infty$ to $T = 0$ at $t = 0$ (MCS). More precisely,
we took totally random spin configurations $\bm{S}_{i}$ as the initial spin configuration
and performed Monte Carlo simulations at $T=0$: each spin flip is accepted only if
it lowers the energy of the system.
To check possible dependence on the details of the algorithm used for the spin updates,
we compared results with ones obtained by single flipping instead of the sub-lattice flipping.

We performed $120$ statistically independent runs for $L=144,288$, $90$ ones for $L=576$, $60$ ones for $L=1152$ and took 
averages of all the physical quantities over the runs and also estimated their  mean-squared errors.

\end{itemize}

\section{Equilibrium Monte Carlo simulations}
\label{sec-equilibrium-MC}

\begin{figure*}[t]
\centerline{\includegraphics[width=1.0\columnwidth]{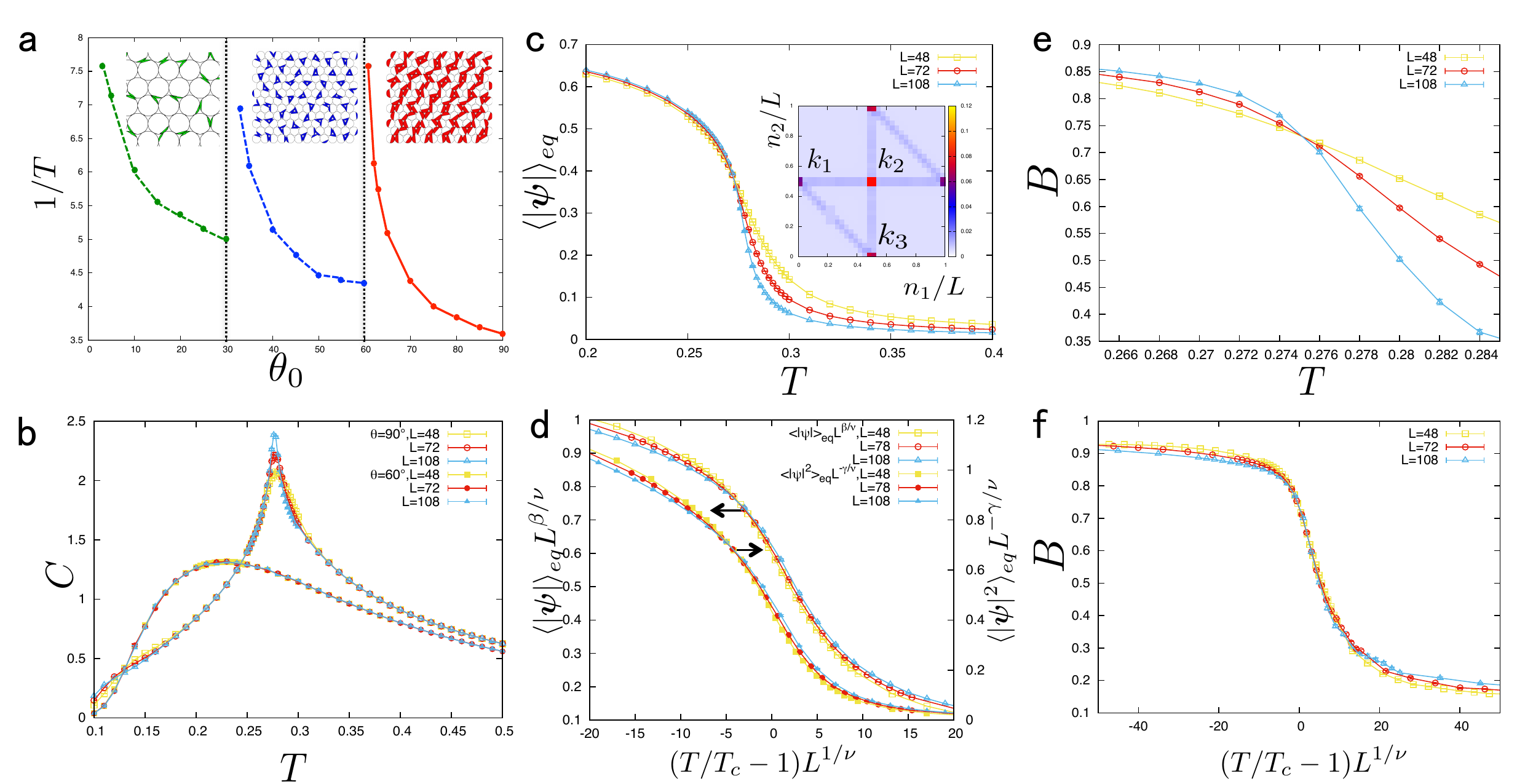} }
\caption{
  Equilibrium properties of the closely packed Janus particles:
  (a) Phase diagram: the solid line (red) is a phase boundary while dashed lines (blue, green) are crossover lines.
  The lines indicate the peak position $T_{\rm peak}(\theta_{0})$ of the heat-capacity (see text for details).
  Representative ground state configurations are also displayed.
  (b) Heat capacity: the data of heat-capacity at $\theta_{0}=60^\circ$ and
  $\theta_{0}=90^\circ$.
  (c) Stripe order parameter: the data of $\langle \psi \rangle_{\rm eq}$  $\theta_{0}=90^\circ$
  is shown. Inset: a density plots of structure factor $S(\bm{k})$ in the reciprocal space at $(\theta_{0}=90^\circ,T=0.1< T_{\rm c})$. 
  (d) Finite size scaling plot of the order parameter  $\langle |\psi|\rangle_{\rm eq}$ and its
  2nd moment $\langle \psi^{2} \rangle_{\rm eq}$ at $(\theta_{0}=90^\circ,T=0.1)$.
  (e) Binder parameter (f) Finite size scaling plot of the Binder parameter.
}
\label{fig_equilibrium_data}
\end{figure*}

\subsection{Phase diagram}

First let us examine the phase diagram. Closely packed Janus particles exhibit three different types of ground states: dimer $0^\circ <  \theta_0 \leq 30^\circ$, trimer $30^\circ <  \theta_0 \leq 60^\circ$ and stripe states
$60^\circ <  \theta_0 \leq 90^\circ$ \cite{shin2014theory,iwashita2014orientational}.
It has been proposed theoretically that there are three low temperature phases corresponding to the
three distinct types of ground states. In the theory \cite{shin2014theory} the dimer and trimer phases are considered
as liquid phases while the stripe phase is considered as a crystalline phase. 
Thus the 1st basic issue to clarify is actual existence of such
thermodynamic phases at low temperatures. We must note however that in the present paper we are not taking into account the vibrational motions of the particles which is included in the theory \cite{shin2014theory}. We assume that the
vibrational motions do not significantly change the nature of the phases.

In Fig.~\ref{fig_equilibrium_data} (a) we show the phase diagram we determined 
based on the measurement of the heat-capacity as we explain below.
There we also show the spatial patterns of the ground states mentioned above.
We define the heat capacity $C$ as,
\beq
C/Nk_{\rm B}= (\beta^{2}/N)(\langle E^{2} \rangle_{\rm eq} -\langle E \rangle_{\rm eq}^{2})
\eeq
where we used the inverse temperature $\beta=1/k_{\rm B}T$. Here after we call $C/Nk_{\rm B}$ simply as $C$.
We measured it by performing equilibrium Monte Carlo simulations at various patch sizes $\theta_{0}$
and temperatures $T$ (more precisely $\tilde{T}=k_{\rm B}T/\epsilon$).
At each values of $\theta_{0}$ we find a peak of the heat capacity at a temperature $T_{\rm peak}(\theta_{0})$ as shown in Fig.\ref{fig_equilibrium_data} (b).
It can be seen that $T_{\rm peak}(\theta_{0})$ slightly depend on the
system size $L$ but only very weakly.
In the panel (a) we display the line $1/T_{\rm peak}(\theta_{0})$  vs $\theta_{0}$ obtained in the system of size $L=72$.
The qualitative feature of the peak temperature $T_{\rm peak}(\theta_{0})$
appears indeed similar to the phase boundary
between the high temperature plastic phase and low temperature phases
proposed theoretically \cite{shin2014theory} (see Fig. 7 (a),(b) of the latter reference), which was observed also by a numerical analysis of the 
spatial patterns of the configurations \cite{iwashita2014orientational} (See Fig. 10 (a) of the latter reference).

In the range $0^{\circ} < \theta_{0} \leq 60^{\circ}$ we found no singularity at $T_{\rm peak}(\theta_{0})$ but just a broad peak
as shown in Fig.~\ref{fig_equilibrium_data} b). Furthermore we found are no appreciable system size dependencies there.
These results suggest $T_{\rm peak}(\theta_{0})$ should be regarded just as a crossover line and that
the dimer and trimer phases are not true thermodynamics phases.
On the contrarily, in the range  $\theta_{0} > 60^\circ$ we find a singular behavior of the heat-capacity at $T_{\rm peak}(\theta_{0})$ as shown in panel b) suggesting a thermodynamic phase transition.

\subsection{Stripe order}

We have seen  that the heat-capacity exhibits a singular behavior suggesting
a thermodynamic phase transition for $\theta_0  >  60^\circ$ into the stripe phase.
Then an important question is whether it is a 1st order transition as suggested
theoretically \cite{shin2014theory} or a continuous transition.
In order to examine the nature of the low temperature phase and the phase transition
we performed further analysis as the following.
In the following we only display results for $\theta_0=90^{\circ}$ for clarity but similar results were
obtained at other values of $\theta_{0} ( > 60^\circ)$ as well.

In order to characterize the low temperature phase, we examined the structure factor defined as,
\begin{eqnarray}
&& S(\bm{k})=\frac{1}{N} \left| \sum_{i=1}^{N} e^{i\bm{r}_i \cdot \bm{k}} \bm{S}(\bm{0}) \cdot \bm{S}(\bm{r}_i) \right|  , \nonumber \\
&& \bm{k}=\frac{n_1}{L}\bm{b}_1+\frac{n_2}{L}\bm{b}_2~(n_1,n_2=0,1,2,...,L-1) ,
\end{eqnarray}
where $\bm{k}$ is a wave vector and $\bm{b}_1=2\pi(1,1/\sqrt{3}),\bm{b}_2=2\pi(0,2/\sqrt{3})$ are reciprocal vectors of the triangular lattice. 
We display in the inset of Fig. ~\ref{fig_equilibrium_data} c) an intensity plot of the structure factor which reveals prominent three distinct spots at $\bm{k}_1, \bm{k}_2, \bm{k}_3$ ($(n_1/L,n_2/L) = (1/2,0),(0,1/2),(1/2,1/2)$
indicating the three possible orientations of the stripes.
%(See Fig.~\ref{fig_stripe}) 

If a stripe structure exhibits a long-ranged order in the low temperature phase, it means
breaking of the 3-fold rotational symmetry.
In order to detect the possible spontaneous symmetry breaking,
let us introduce an order parameter $\bm{\psi} = (\psi_x,\psi_y)$ defined by,
\begin{eqnarray}
&& \psi_x=\frac{\sqrt{3}}{2}(S(\bm{k}_2)-S(\bm{k}_3)), \nonumber \\
&& \psi_y=S(\bm{k}_1)-\frac{1}{2}(S(\bm{k}_2)+S(\bm{k}_3)).
\end{eqnarray}
Clearly, it will be averaged out to $0$ in a thermal average
if the $3$-fold rotational symmetry is preserved.
In Fig.~\ref{fig_equilibrium_data} (c) we show the temperature dependence of the absolute value
of the order parameter $\langle |\bm{\psi}| \rangle_{\rm eq}$.
It can be seen that it rapidly increases in the vicinity of $T_{\rm peak}(\theta_{0}=90^{\circ})$ by lowering the temperature strongly suggesting spontaneous breaking of the 3-fold rotational symmetry.

To analyze the phase transition more precisely, we examined the Binder's parameter defined by,
\beq
B = \frac{3}{2}\left (1-\frac{\langle \psi^4 \rangle_{\rm eq}}{3\langle \psi^2 \rangle_{\rm eq}^2}\right).
\eeq
As shown in Fig.~\ref{fig_equilibrium_data} (e), the data of different system sizes exhibit a crossing feature
strongly suggesting a continuous phase transition.
From the latter we estimate the critical temperature of the system as $T_c\approx0.275$ at $\theta_{0}=90^{\circ}$, which is in good agreement with the peak temperature $T_{\rm peak}(\theta_{0}=90^{\circ})$ of the heat-capacity (See Fig.~\ref{fig_equilibrium_data}).

Finally, to examine the universality class of the continuous phase transition,
we performed finite size scaling analyses as follows.
Since the order parameter has the $3$-fold rotational symmetry, it is natural to expect that the universality class of the
transition is the same as that of  the 3-state Potts model in 2d, whose critical exponents are known to be
$\nu=5/6,\beta = 1/9,\gamma=13/9$ \cite{wu1982potts}.
In Fig.~\ref{fig_equilibrium_data} (d) and (f) we display finite size scaling plots of the order parameter
$\langle |\bm{\psi}| \rangle_{\rm eq}$, its 2nd moment $\langle \bm{\psi}^{2} \rangle_{\rm eq}$
and the Binder's parameter $B$. The successful collapse of the data in the scaling plots strongly suggests
that the universality class of the phase transition into the stripe phase
is indeed that of the 3-state Potts model in 2d. Thus our result is qualitatively different from that of the the harmonic theory \cite{shin2014theory} which proposes a 1st order phase transition.

\section{Out-of equilibrium Monte Carlo simulations}
\label{sec-out-of-equilibrium-MC}

\begin{figure}[t]
\centerline{\includegraphics[width=0.8\columnwidth]{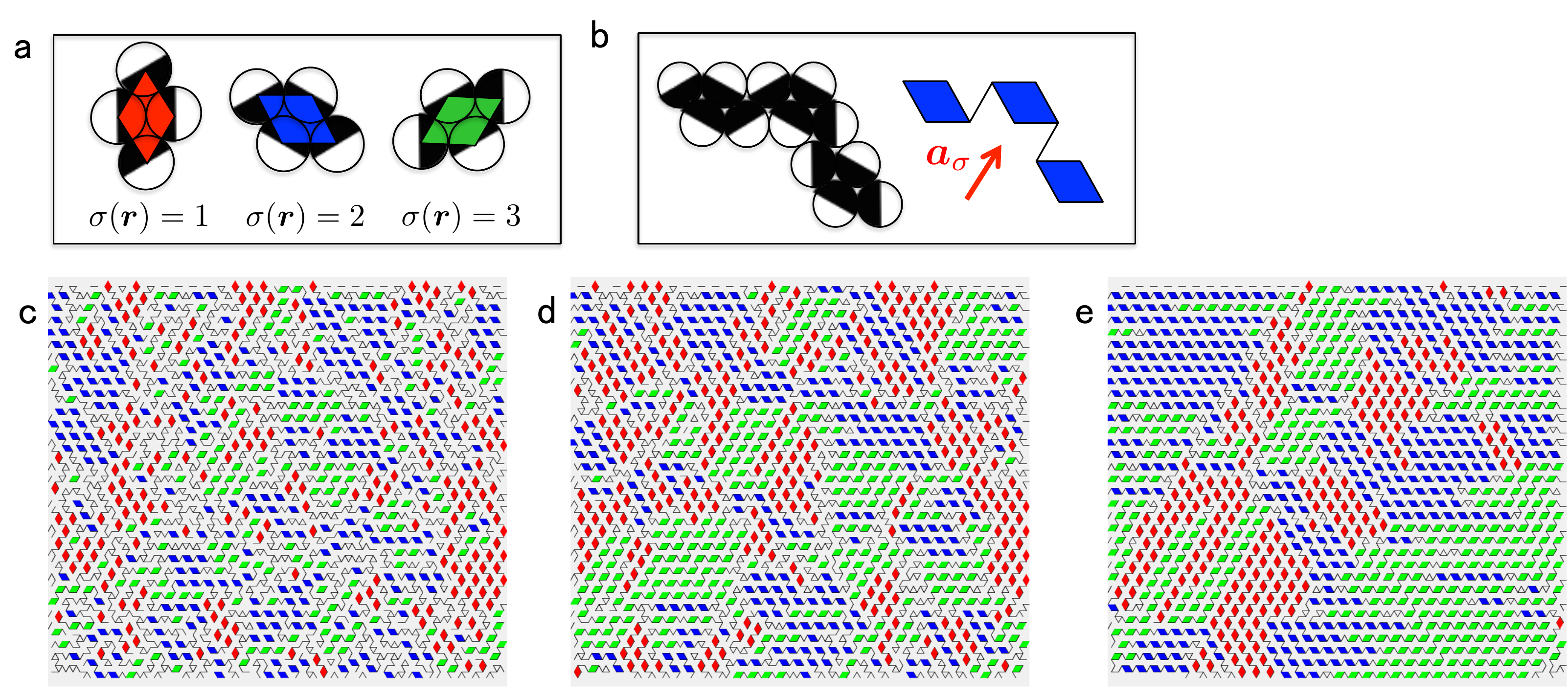} }
\caption{
  Spatial organization of the orientational degree of freedom in the stripe phase. (a) {\bf Tetramers:}  Stripes are composed of tetramers, each of which can be represented by a {\it rhombus}.
   There are three possible orientations of the tetramers (rhombus) $\sigma=1,2,3$
   due to the 3-fold rotational symmetry of the triangular lattice.   
   (b) {\bf Stripes:} A stripe is a train of tetramers with the same
   orientation $\sigma$. (Here the case  $\sigma=2$ is shown as an example.)
   Such a train can be continued indefinitely without any energy cost
   along two directions among $\bm{a}_{1}$,$\bm{a}_{2}$,$\bm{a}_{3}$
   except for $\bm{a}_{\sigma}$.
(c)-(e) {\bf Snapshots} 
These are snapshots of the system during a Monte Carlo run
of a system with $L=64$ at times (c) $t=80$ (d) $t=320$ and (e) $t=1280$ after a temperature quench from $T = \infty$ to $T=0$. The colored rhombus represent the oriented tetramers.
Note that there are black lines representing attractive contacts which do not belong to the tetramers, whose number progressively decreases with time.
}
\label{fig_stripe}
\end{figure}

In previous section, we demonstrated that there is a continuous phase transition into the stripe phase for
patch size in the range $60^\circ < \theta_0 \le 90^\circ$.
Now we turn to study the relaxational dynamics after temperature quenches into the stripe phase. Most natural expectation would be that the relaxation of the system proceeds as a coarsening process \cite{bray2002theory} of the stripes. However an experimental study \cite{jiang2014orientationally} has suggested
some sort of glassy dynamics.
%In order to 
%clarify the nature of the relaxational dynamics in this section.

\subsection{Local order parameter}

By inspecting the stripe structure of the ground state (see Fig.~\ref{fig_equilibrium_data} (a)),
we first notice that a stripe consists of a train of tetramers.
In a tetramer, four particles are in attractive contacts with each other.
A tetramer can be represented by a rhombus whose ridges connect the
centers of the particles in the tetramer as shown in Fig.~\ref{fig_stripe} (a).
There are three possible orientations of the tetramers (rhombus)
due to the 3-fold rotational symmetry of the triangular lattice.   
We label them by three colors: red, blue and green
or the corresponding integers $\sigma=1,2,3$.
The latter integers corresponds to vectors $\bf{a}_{\sigma}$
($\sigma=1,2,3$)(See Fig.~\ref{fig_janus}), each of which is  parallel to the
vector connecting the shorter diagonal of the corresponding rhombus.

A stripe is a train of tetramers with the same orientation (color)
$\sigma$ as shown in Fig.~\ref{fig_stripe} (b). 
In such a train, two adjacent tetramers are connected by a link
parallel to $\bm{a}_{\sigma}$,  along which two particles in different tetramers are in an attractive contact.
Such a train can be continued indefinitely without any energy cost
along two directions among $\bm{a}_{1}$,$\bm{a}_{2}$,$\bm{a}_{3}$
except for $\bm{a}_{\sigma}$.

We performed out-of equilibrium Monte Carlo simulations starting from $T=\infty$ random initial
configurations. Since spin updates are performed at $T=0$, the system simply tries to lower
the energy increasing the number of attractive contacts.
In Fig.~\ref{fig_stripe} (c)-(e) we show some snapshots of the configuration of the system
represented by the oriented (colored) rhombuses. There we also display the attractive contacts which
are not included in the tetramers as black lines. It can be seen that the length of stripes, namely
trains of tetramers with the same orientations, increase with time.

Moreover we find that stripes with the same color also grow into their transverse directions
resulting in growth of 2 dimensional domains of tetramers with the same color.
This is not so obvious at a first sight because
stripes of different colors can run in parallel in a ground state.
For instance, stripes of $\sigma=1$ and $\sigma=2$ can be extended in parallel along
the same direction $a_{3}$. In other words, there are large number of ground states
in which all stripes are directed toward a given direction $a_{\sigma}$ $\sigma \in (1,2,3)$
where each of the stripes can take one of the 2 possible values $\sigma'( \neq \sigma)$ arbitrarily.
Indeed we can finds some stripes of different colors running in parallel
in the snapshots shown in Fig.~\ref{fig_stripe}.

Where stripes of different colors running into different directions meet, 
we find a local region not belonging to tetramers. There the local energy is higher as compared with
the bulk of the domains. Thus the latter may be regarded as energetic defects.
Energy relaxation should take place by reducing the area of such energetic defects
which amount to expand some domains at the expense of the others, i.~e. domain growth.
Note also that a stripe can be bended without any energy cost as shown in Fig.~\ref{fig_stripe} (b).
Indeed we can find such configurations in the bulk of the domains. Presumably this bending
mechanism allows formation of 2 dimensional domains made of stripes which are 1 dimensional.

Based on the above observation let us introduce a local order parameter as follows.
We define a Potts variable  $\sigma_i(t)$ at each site $i=1,2,...,N$ such that
it equals to the color $\sigma \in (1,2,3)$ of the tetramer (rhombus) it belongs to at time $t$.
If it does not belong to any tetramers, we assign $\sigma_{i}(t)=0$.

\subsection{Dynamical observable}

\begin{figure*}[t]
\centerline{\includegraphics[width=1.0\columnwidth]{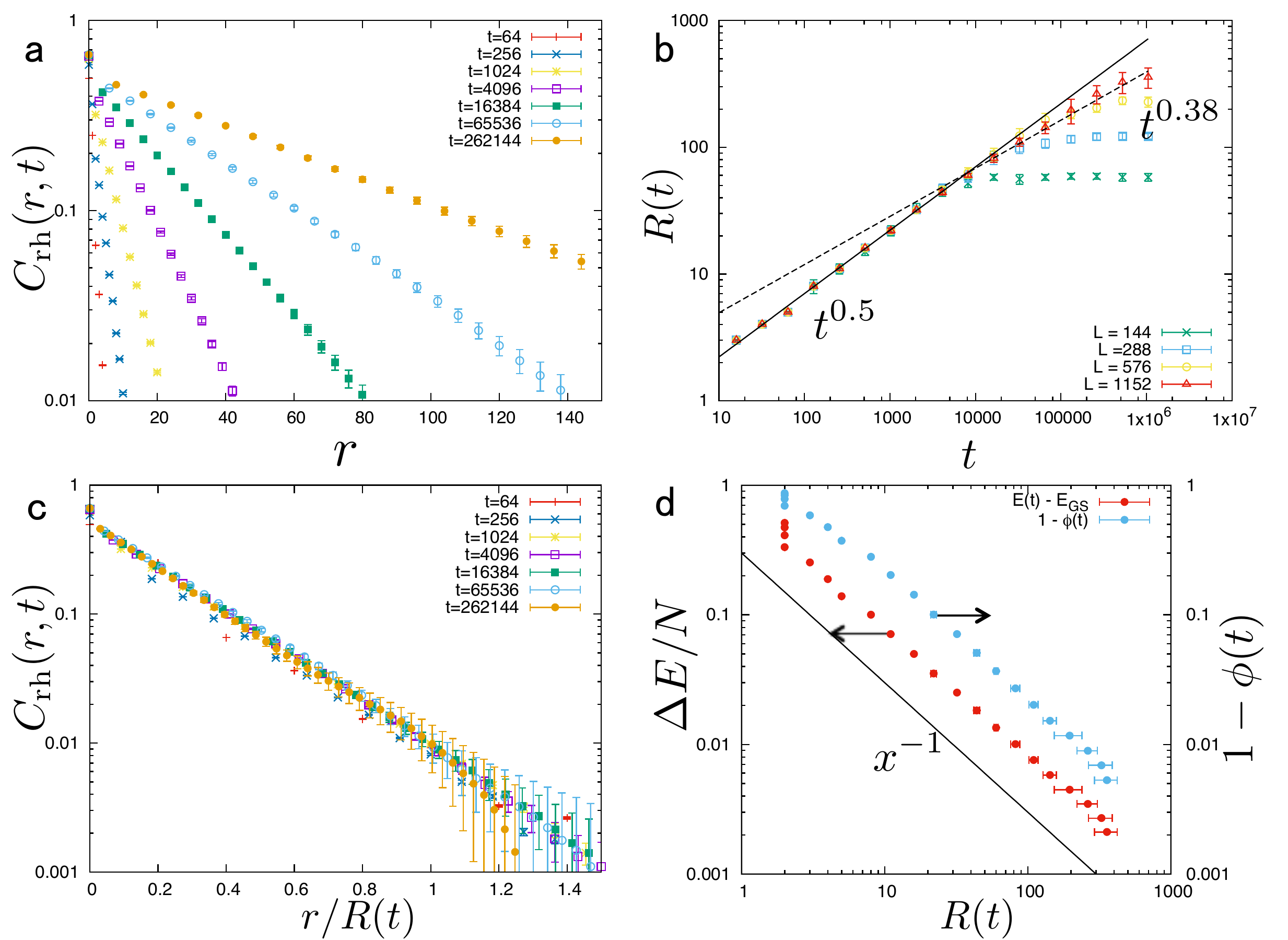}} 
\caption{
Domain growth and relaxation after temperature quenches.
(a) Rhombus spatial correlation function $C_{\rm rh}(r,t)$.
(b) Time dependence of the domain size $R(t)$ (see text for the definition).
(c) Scaling plot of the rhombus correlation function
$C_{\rm rh}(r,t)$ in terms of the domain size $R(t)$ assuming the scaling form \eq{scaling_spatial}.
(d) Relaxation of energy density and rhombus density.
All data presented here (except for (b)) were obtained in systems with $L=1152$. 
}
\label{fig_relaxation}
\end{figure*}

Now let us introduce several dynamical quantities to study the relaxational dynamics based on the local order parameter $\sigma_{i}(t)$ introduced above.
First, we introduce the density  $\phi(t)$ of the rhombuses by,
\begin{equation}
  \phi(t) = 1 - \frac{1}{N}\sum_{i=1}^N \delta_{\sigma_{i}(t),0}.
  \label{eq-rh-density}
\end{equation}
Note that $1-\phi(t)$ is the density of spins not belonging to the tetramers. We also measure the relaxation
of energy density $E(t)/N$.

To characterize the domain growth, a spatial correlation function of the Potts variable  can be introduced as,
\beq
C_{\rm rh}(r,t)=P(r,t) - \lim_{r^\prime \to \infty}  P(r^\prime,t) 
\eeq
with
\begin{eqnarray}
&&    P(r,t) = \frac{\sum_{i,j=1}^{N} \delta_{r,r_{ij}}P_{ij}(t)}{
    \sum_{i,j=1}^{N} \delta_{r,r_{ij}}}
  \qquad 
  P_{ij}(t)=\delta_{\sigma_{i}(t),\sigma_{j}(t)}
(1-\delta_{\sigma_{i}(t),0}\delta_{\sigma_{i}(t),0}),
\end{eqnarray}
where
$\delta_{\alpha,\beta}$ is Kronecker's delta. 
In simulations, since we can't take a limit $\lim_{\bm{r}^\prime \to \infty}$, we subtract correlation with the most distant particles, i.~e. $r_{ij}=L/2$. 

%In the following we present the data obtained with $L=1152$
%and comment on finite size effects when necessary.

\subsection{Domain growth}

Now let us discuss our results on the dynamics. 
In Fig.~\ref{fig_relaxation} (a) we display the rhombus spatial correlation function. 
We can see that the correlation length becomes larger with time reflecting the domain growth.
We evaluated the domain size $R(t)$ as the shortest distance $r$ such that $C_{\rm rh}(r,t) < 0.01$.
As shown Fig.~\ref{fig_relaxation} (b), we find that the resultant domain size grows initially as
$R(t) \sim t^{0.5}$. However it crossovers to slower growth law $R(t) \sim t^{0.38}$ at larger time scales
$t^\ast \approx 10^4$.
The data suggest this crossover is not a finite size effect. 
%Specifically, we obtained the same result with  $L = 1152,576$ in the same time scales
%while some finite size effects can be seen in a smaller systems.

The initial growth law $R(t) \sim \sqrt{t}$ is the same as in
the standard coarsening process of non-conserved order parameters \cite{bray2002theory}.
On the other hand, the slower growth law found at larger time scales is not a standard one
and deserves to be studied further in the future.

In Fig.~\ref{fig_relaxation} (c) we show a scaling plot of the rhombus spatial correlation function 
assuming a scaling form,
\begin{equation}
C_{\rm rh}(r,t)=\tilde{C}\left( \frac{r}{R(t)} \right).
\label{scaling_spatial}
\end{equation}
It is clear that the scaling law is satisfied as in standard coarsening systems \cite{bray2002theory}.
Remarkably the scaling function is universal in spite of the fact that the growth law $R(t)$ itself
exhibits the crossover behavior mentioned above. Such robustness of the scaling function
of the spatial correlation function is known also in some other cases \cite{bray2002theory}.

\subsection{Energy and rhombus density relaxation}

Next let us examine the relaxation of the energy and the rhombus density.
Since we are working at $T=0$, the energy is expected to converge to that of the ground state
which is $E_{\rm GS}=-1.5N$. A natural scaling ansatz for the relaxation of the energy is 
\begin{equation}
\Delta E = E - E_{\rm GS} \sim R(t)^{d_{s} - d} \qquad {d_{\rm s}-d=-1}
\end{equation}
where $d$ and $d_{\rm s}$ are spatial dimension of the system and fractal dimension of domain wall respectively.
The last equation $d_{s}=d-1$ holds if the domain is compact as in usual coarsening systems. 
In Fig.~\ref{fig_relaxation}(d) we show the relaxation of the energy density $\Delta E(t)/N$.
This result is consistent with the expected scaling law. 
We checked that the scaling behavior does not change by using single-spin flipping instead of the sub-lattice flipping.

Similarly we expect that the density of the rhombus density behaves as,
\begin{equation}
1-\phi(t)  \sim   R(t)^{d_{s} - d} \qquad {d_{\rm s}-d=-1},
\end{equation}
because the density $1-\phi(t)$ which do not belong to the rhombus should correspond to the area
where the local energy is high, i.~e. domain wall.  As shown Fig.~\ref{fig_relaxation} (d)
this scaling law is also satisfied.

Again, it is remarkable that the scaling behaviors in terms of the domain size $R(t)$ are universal in spite of the fact that the growth law $R(t)$ itself exhibits the crossover behavior as mentioned before.

\subsection{Two time autocorrelation function}

\begin{figure*}
\centerline{\includegraphics[width=1.0\columnwidth]{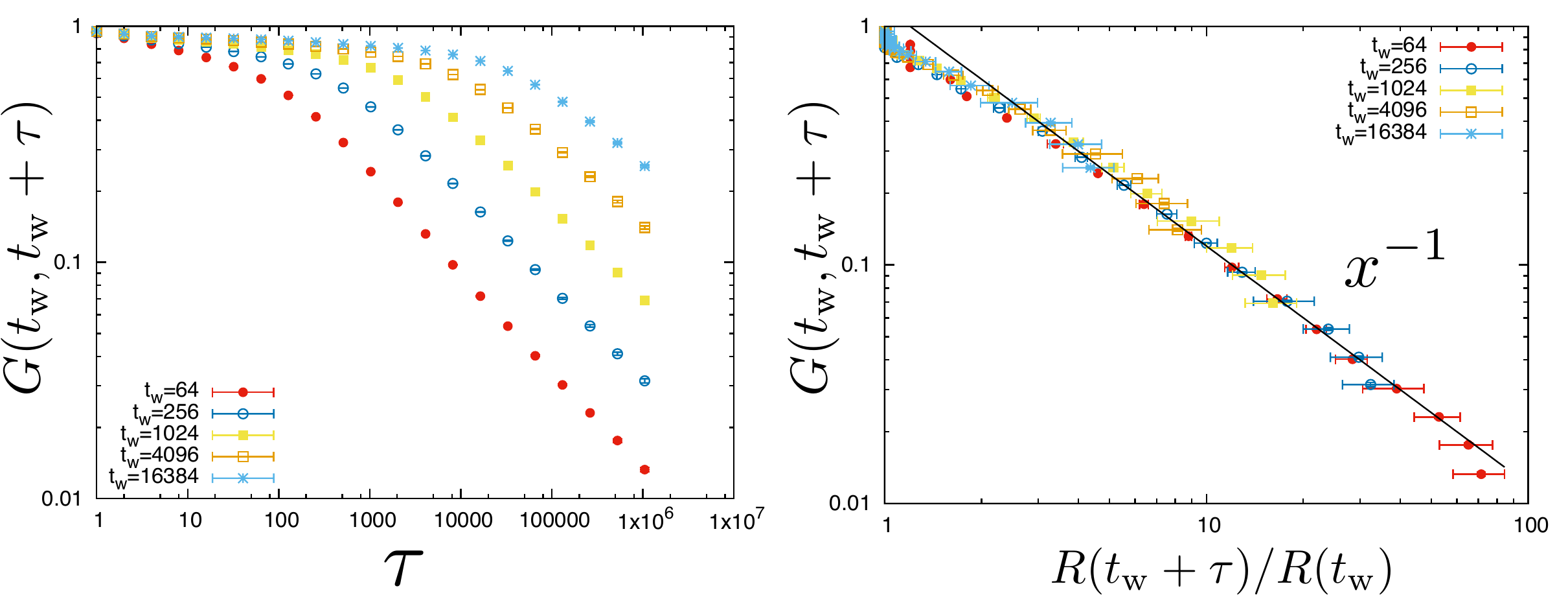} }
\caption{
The spin auto-correlation function: 
(a) The waiting time $t_{\rm w}$ dependence of the spin auto-correlation function plotted
against the time difference $\tau = t - t_{\rm w}$.
(b) Scaling plot of the spin auto-correlation function.
All data presented here were obtained in systems with $L=1152$. 
}
\label{fig_aging}
\end{figure*}

Finally let us discuss the spin-auto correlation function,
\beq
C(t,t_{\rm w})=\frac{1}{N}\sum_{i=1}^{N}\bm{S}_i(t)\cdot\bm{S}_i(t_{\rm w}).
\label{eq-spin-autocorrelation-function}
\eeq
As shown in Fig.~\ref{fig_aging} (a), the two time quantity strongly depends on the waiting time
$t_{\rm w}$ as expected.
A natural scaling ansatz is \cite{fisher1988nonequilibrium,bray2002theory,newman1990dynamic},
\begin{equation}
C(t,t_{\rm w})=f\left(\frac{R(t)}{R(t_{\rm w})}\right)\sim\left(\frac{R(t)}{R(t_{\rm w})}\right)^{-\lambda}~~(t \gg t_{\rm w}).
\label{scaling_auto}
\end{equation}
It is clearly satisfied by our data as shown in Fig.~\ref{fig_aging} (b).
The result suggests that the non-equilibrium dynamical exponent $\lambda=1$.
Once again, it is remarkable that the scaling behavior in terms of the domain size $R(t)$ is universal in spite of the fact that the growth law $R(t)$ itself exhibits the crossover behavior as mentioned before.

\section{Implications for experiments}
\label{sec-implications-for-experiments}

Let us discuss implications of our results for experiments.
In equilibrium simulations discussed in sec. ~\ref{sec-equilibrium-MC}, we made observations
by slowly cooling the system (decreasing $k_BT/\epsilon$) keeping the system in equilibrium.
In experiments of colloidal particles, it may be possible to realize quasi-static, slow changes of the strength of effective interaction $\epsilon$
via changing the properties of the solution \cite{iwashita2013stable}.
The structure factor $S(\bm{k})$ is well accessible in experiments \cite{iwashita2014orientational}.
Thus it would be very interesting to perform systematic analysis of the nature of the possible phase transitions
in experiments and make comparisons with our results.
We have to note however that the effects of the gravity, friction between the particles and friction between the particle and substrate which may play some roles in experiments are not included in our present model.

The dynamics of the stripes was investigated in a previous work \cite{jiang2014orientationally} 
both experimentally and numerically.
The spin auto-correlation function \eq{eq-spin-autocorrelation-function} was also studied
but unfortunately without paying attention to the waiting time $t_{\rm w}$ effect.
There the decay of the spin auto-correlation function was fitted to 
a stretched exponential function $A\exp(-Bt^\beta)$ which was interpreted as a signature of
glassy dynamics. Indeed our data shown in Fig.~ \ref{fig_aging} can also be fitted  by such
a formula. However the result of our scaling analysis in terms of the domain size  $R(t)$
presented in the previous section strongly
suggests that the relaxational dynamics, including the aging effect, can be simply understood as a consequence
of a coarsening process. Moreover we consider that it is rather unlikely to realize
super-cooled, glassy liquid state since it turned out that the transition into the stripe phase
is a 2nd order transition rather than a 1st order transition.

Comparing our data with the result presented in  Ref \cite{jiang2014orientationally}, we expect our
1(MCS) roughly corresponds to  $10^{-2}$(sec) in the experiment. Thus we expect the experimental time scale
$O(10^{3})$ (sec) is large enough to investigate the domain growth process we reported.
However finite size effects should be severe in the previous experiment which uses $L\approx50$.
Indeed we found strong finite size effects for smaller system sizes as shown in Fig.~\ref{fig_relaxation} (b).

\section{Conclusion}
\label{sec-Conclusion}

To summarize, we studied the equilibrium and dynamic properties of directors of the closely packed Janus particles
by Monte Carlo simulations. Although the gross phase behavior we found is in agreement with the previous
theoretical\cite{shin2014theory} and experimental works \cite{iwashita2014orientational},
the present work clarified the precise nature of the phase transitions.
In equilibrium, we found that the transitions from the high temperature plastic phase to low temperature dimer ($0^\circ < \theta_0 \le 30^\circ$) and trimer ($30^\circ < \theta_0 \le 60^\circ$)
'phases' are smooth crossovers. On the other hand, in $60^\circ < \theta_0 \le 90^\circ$
we found 2nd order phase transition into stripe phase whose universality class turned out to be the same as that of the
$3$-state Potts model in $2$-dimensions. 

In relaxation dynamics in the stripe phase $60^\circ < \theta_0 \le 90^\circ$ bellow $T_{\rm c}$, we found strong evidences
of coarsening process.  The growth law of the domain size $R(t)$ exibits a crossover
from the usual behavior  $R(t) \sim \sqrt{t}$ at shorter time scales to a slower growth at longer time scales.
At the moment the mechanism of the crossover is unknown.  
Nevertheless scaling analysis of various physical observable in terms of the domain size $R(t)$
reveal salient universal features of the coarsening process of stripes.
Such robustness of the scaling features in terms of the domain size with some complicated crossover behaviors
renormalized into the domain growth law is known in some systems \cite{yoshino2002extended,corberi2011growing}.
We found no evidence of glassy dynamics originally anticipated in \cite{jiang2014orientationally}.
However we believe the possibility to realize orientational glasses
using patchy colloids is a legitimate and interesting problem by itself
\cite{yoshino2017rotational}
which should be explored more extensively.

Finally we note that we have not considered at all possible effects of translational motions of the particles
in the present paper. In 2 dimensions translational thermal fluctuations can be strong  \cite{mermin1966absence}.
Also coupling between translational and rotational degree of freedom should bring about more varieties of
phenomena. Certainly these points should be examined in the future.

\begin{acknowledgments}
The author thanks Yuliang Jin, Macoto Kikuchi and Takenobu Nakamura for useful discussions. This work was supported by KAKENHI (No. 25103005  ``Fluctuation \& Structure'' and No. 50335337) from MEXT, Japan. 
This work was achieved through the use of OCTOPUS and supercomputer system SX-ACE
at the Cybermedia Center, Osaka University.
\end{acknowledgments}

\bibliography{janus}

\end{document}